\documentclass[sn-basic,Numbered,iicol,pdflatex]{sn-jnl}

\usepackage[T1]{fontenc}   
\usepackage{lmodern}       
\usepackage{amsmath}
\usepackage{amssymb}
\usepackage{amsfonts}
\usepackage{xcolor}
\usepackage{textcomp}
\usepackage{manyfoot}
\usepackage{booktabs}
\usepackage{array}
\usepackage{tabularx}
\usepackage{pifont}
\usepackage{enumitem}
\usepackage{placeins}
\usepackage{microtype}

\newcolumntype{R}[1]{>{\raggedright\arraybackslash\hspace{0pt}}p{#1}}
\newcolumntype{Y}[1]{>{\hsize=#1\hsize\raggedright\arraybackslash\hspace{0pt}}X}
\newcolumntype{Z}{>{\raggedright\arraybackslash\hspace{0pt}}X}

\begin{document}

\title[Classifying Attachments and Affective Social Bonds in HRI]{A Procedure for Classifying Attachments and Affective Social Bonds in Human--Robot Dyads}

\author*[1]{\fnm{Imran} \sur{Khan}}\email{imran.khan.3@warwick.ac.uk }

\author[2]{\fnm{Elisabeth} \sur{Blagrove}}\email{e.blagrove@warwick.ac.uk}

\affil*[1]{\orgdiv{Department of Computer Science}, \orgname{University of Warwick}, \orgaddress{\city{Coventry}, \postcode{CV4 7AL}, \country{United Kingdom}}}

\affil[2]{\orgdiv{Department of Psychology}, \orgname{University of Warwick}, \orgaddress{\city{Coventry}, \postcode{CV4 7AL}, \country{United Kingdom}}}

\abstract{Human--robot interaction (HRI) claims that people form \emph{attachments} and
\emph{social bonds} with artificial agents, yet the terms are often applied without the
behavioural and physiological criteria that give them content in their source disciplines.
Without this empirical grounding, studies deploy widely divergent methods, frequently producing expansive relational claims that far outstrip their underlying evidence. To address this, we propose a standardised four-question procedure, grounded in criteria established in the developmental, ethological, and neuroendocrine literatures, that classifies a given human--robot tie as an \emph{attachment}, an \emph{affective social bond}, or \emph{no relationship}, with intermediate classifications when evidence is
incomplete. We specify minimum evidential requirements for each question, and provide
candidate HRI study designs, adapted from validated human--human, human--animal, and
animal--animal paradigms. We then demonstrate the procedure by applying it to a
representative set of published HRI studies, showing how often relational claims outstrip
what the reported designs can establish. Finally, we discuss the ethical and regulatory
burdens created when artificial agents engage human biobehavioural systems. By replacing the
divergent operationalisations with a unified, criterion-based classification, this paper
gives HRI practitioners a standardised basis for evaluating, classifying, and comparing
human--robot relationships, and sets out the experimental rigour that each classification
demands. We therefore call on researchers of human--robot relationships to adopt such
rigour, or to consider alternative terminology in their descriptions of these ties.}


\maketitle

\section{Introduction}
\label{sec:intro}
A developmental psychologist, a behavioural neuroscientist, and a roboticist walk into a bar. The bartender asks each of them to write down how they define \emph{attachment}. The developmental psychologist may write about a behavioural system for regulating felt security \cite{sroufe1977attachment}, engaged by threat, present when a particular figure is sought for comfort under distress and used as a base from which to explore when distress is absent \cite{bowlby1969,ainsworth1978}. In contrast, the behavioural neuroscientist may allude to an enduring preference for a familiar partner, sustained by specific neuroendocrine mechanisms, and established when that partner is chosen over an unfamiliar alternative \cite{carter1998}. The roboticist, looking at the current evidence from their field, may conclude that it indicates a human has come to like \cite{brandtzaeg2022}, trust \cite{belpaeme2012}, or report positive feelings towards \cite{mitchell2025} a machine, or likes its appearance \cite{eum2026}, and would be sorry to lose it.

Prima facie, these responses cannot all be naming the same concept. The developmental psychologist offers canonical definition of attachment, while the neuroscientist describes something closer to a social bond; our story's roboticist is, regrettably, describing neither. The story would be similar should our inquisitive bartender ask them about \emph{social bonding}, with the roboticist concluding a bond means improved rapport \cite{diazboladeras2023}, trust \cite{salekshahrezaie2021}, or sustained engagement with an agent over time \cite{tanaka2007socialization,barakova2018}.

This confusion between the terms is not new \cite{carter2005attachment}, and previously, could be considered productive \cite{feldman2017}. However. in human--robot interaction, this confusion has become critical, because the field now designs for \cite{eum2026,szondy2024attachment, yang2013}, measures \cite{cheng2026,lonklang2025robot}, and makes claims about \cite{konijn2025,you2017emotional, khan2026} human--robot relationships whose quality it cannot characterise accurately \cite{mitchell2025,rabb2022}. HRI has adopted the vocabulary of close relationships---``attachment'' and ``bonding''---without importing the criteria that make that vocabulary meaningful. The result is that changes in interaction quality \cite{tanaka2007socialization},  reported attitude \cite{cheng2026}, or engagement \cite{barakova2018} can each be given the same relational label, despite having no shared qualia. More importantly, a reader has no way to tell which of them was actually observed. Therefore, a researcher studying human--agent relationships cannot proceed to questions about the relationship's dynamics---its development over different timescales, effects on participant behaviours and experiences under design manipulation, effect magnitude---because those questions presuppose an answer to a more fundamental one: what kind of relationship is under investigation in the first place?

Attachment, typically invoking attachment theory \cite{bowlby1969}, is both mischaracterised and overgeneralised in HRI studies \cite{mitchell2025}. The tests that identify attachment in the \citet{bowlby1969} and Ainsworth \cite{ainsworth1978} tradition are readily available, standardised, and already appled across a number of species, and the species boundary \cite{palmer2008,horn2013securebase,gacsi2013safehaven,mariti2013owners}. However, they are largely absent from human--robot relationship studies, oftentimes, besoke self-report instruments have been applied \cite{mitchell2025,cheng2026,lonklang2025robot}. Likewise, ``social bonding'' in HRI has been reported via adjacent measures, such as a desire for future interaction \cite{xu2025care}, labelling a robot as a ``friend'' or ``classmate'' \cite{belpaeme2012},  sustained or increased engagement \cite{tanaka2007socialization,barakova2018}, and self-disclosure and warmth \cite{neerincx2021childbonding}. Changes in these measures may, indeed, be constitutive of a social bond, but none is, in isolation, evidence to suggest two agents are socially bonded in the sense construed by the ethological and neuroendocrine literatures.

Despite the growing body of literature using these terms to describe human--robot dyads, we find a marked lack of any consistent, suitable criteria that warrant the claims that a human is either ``attached'' or ``bonded'' with a robot partner. Therefore, we argue that both concepts require more rigorous operationalisation than HRI currently offers. We also suggest that this rigour already exists (and is validated in various disciplines), and thus, needs to be more stringent adoption within the field. Building on developmental psychology, comparative, and neurobiology literatures, we propose that ``attachment'' in HRI should be established behaviourally, through the security functions that define it. Social bonding, a related-yet-distinct system \cite{ettenberger2021, konijn2025}, should be established through selective preference for a particular partner, and identification of affective, physiological, and behavioural regulation (if any) that a particular agent supplies. 

We position this paper amongst five recent papers set out to clarify how HRI identifies such relationships. \citet{rabb2022} import Bowlby's and Ainsworth's features and arrange all possible human--robot ties along a weak-to-strong attachment \emph{spectrum}. We accept the features, but dispute the ordering, for reasons taken up in Section~\ref{sec:magnitude} below. In the same year, \citet{law2022} catalogued the loose use of ``attachment'' in HRI, traced the definition on which attachment is the cumulative sum of a user's emotional episodes with a device \cite{norman2004}, and concluded that the term ``should be reserved for relationships that feature the four different behaviors and functions identified in social psychology''. The review by \citet{mitchell2025} found that instruments measuring attachment in HRI ``do not measure attachment directly but instead assess factors associated with attachment'', while defining human--robot attachment as an ``emotional bond facilitated by prolonged interaction, a unique robot identity, and effective communication''. The authors state the canonical definitions correctly, however, then set them aside deliberately; on the grounds that strict definition would exclude the weaker ties that make up most of HRI. We take up this argument in Section~\ref{sec:magnitude}. \citet{konijn2025} have recently proposed a Theory of Affective Bonding for human--robot relationships, define bonding as \textit{``feeling emotionally or affectively connected, or social affection toward someone''}. Here, they step away from attachment theory altogether:\textit{ ``in avoiding this kind of connotations, we argue that the concept of bonding is more appropriate to describe human--robot relationship formation as a process''}. We support this framing (discussed more in Section \ref{sec:bonding}), %
and, therefore, refer to this type of relationship as an ``affective social bond'', referring to a tie in which a \emph{particular} partner regulates another's affective and physiological state. Most recently, \citet{banks2026mc} review 71 works on \emph{machine companionship} (one of the terms \citet{mitchell2025} pooled at retrieval) published between 2017 and 2025, and found more than fifty distinct measured variables among the corpus. They offer a synthesised definition of companionship as \textit{``an autotelic, coordinated connection between human and machine that unfolds over time and is subjectively positive'}'. The same pattern of divergent metrics recurs across integrative bonding theories, ``synthetic relationship'' (i.e. human--AI relationships \cite{defreitas2026hyper,lee2026design}), staged process models \cite{chen2026,fabes2026}, or mind-perception attributions \cite{ventura2026,rupprechter2026,shu2026haia,boyd2026mira}

Taken together, this work describes a single problem from different directions: that no current human--robot relationship qualifies as an attachment \cite{law2022}, there is little evidence of strong attachment in the field \cite{rabb2022}, the instruments do not measure what they name \cite{mitchell2025}, measured variables do not have an agreed construct behind them \cite{banks2026mc}, and the term is better set aside altogether \cite{konijn2025}. This paper agrees with all five verdicts. What is required in this ongoing discussion is a standardised procedure for HRI researchers to apply to their studies, in order to make accurate claims about the specific dyadic relationship under investigation. Such a procedure would allow the field to converge on a shared set of measurements and requirements, against which a given relationship can be qualified as an attachment, as a social bond, or neither.

\subsection{Contributions}

The primary contribution of the present paper is a classification procedure for operationalising human--artificial ties (Section~\ref{sec:operationalising}). The procedure sets out four questions, each answerable through established experimental paradigms. Together, these return a determinate classification of a given human--robot tie as either an \emph{attachment}, an \emph{affective social bond}, or identify where no relationship exists. We also propose three related classifications where criteria are only partially-met (Table~\ref{tab:classify}). For each point in the procedure, we state the minimum empirical evidence it requires to be satisfied, and offer candidate study designs for HRI researchers, adapted from validated paradigms in the human--human, human--animal and animal--animal literatures. 

To ground this procedure and make it actionable for researchers, the paper provides two further enabling contributions. Firstly, we recover operational definitions and evidence criteria for attachment and affective social bonding from developmental, ethological, and neuroendocrine literatures. We offer a short synthesis of existing evidence to ground these criteria, as a means to support researchers in designing empirical studies. To further assist researchers with study design, we compile a targeted inventory (Table~\ref{tab:reg}) mapping biobehavioural regulatory substrates to observable proxies and feasible HRI laboratory instruments. Secondly, we evaluate current HRI definitional practices against source disciplines, providing an argument against reducing complex, multi-dimensional relational ties to scalar magnitudes. Finally, we demonstrate our procedure's utility by applying it to a representative set of existing HRI studies, evaluating reported terminology against what their experimental designs established through our proposed procedure (Section~\ref{sec:apply}). 

Together, these contributions resolve the diagnostic and reporting gaps highlighted across recent HRI literature \cite{law2022, rabb2022, mitchell2025, banks2026mc, konijn2025}, establishing a standardised foundation for conceptualising and operationalising human--robot ties. By replacing proxy and divergent metrics with a criterion-based taxonomy, this framework enables the field to generate cumulative, empirically comparable evidence regarding the precise relational states instantiated in human--robot interaction.

\section{Attachment and Affective Social Bonding as Separate Systems}
\label{sec:home}
Although the two traditions are related \cite{ainsworth1989}, with terms sometimes used interchangeably \cite{carter2005attachment}, ``attachment'' and ``affective social bonding'' denote distinct functional and mechanistic systems, individuated by what they \textit{do} rather than by what they are \textit{called}. \textbf{Attachment} defines what a relationship is functionally \emph{for}: it is a threat-gated system for regulating felt security, specified by \textit{safe-haven} and \textit{secure-base} functions and identified through behavioural responses under threat or challenge. An \textbf{affective social bond}, by contrast, concerns how a relationship operates biobehaviourally: it is an arrangement for modulating affective and physiological states through a \emph{particular} partner. It asks which partner an organism is oriented to, and what that partner does to its body---ongoing (co-)regulation. This is supported by oxytocin--dopamine reward mechanisms---measured by partner-preference tests together with behavioural and physiological measurement. This section assembles the criteria that define these two systems in their originating disciplines, alongside the operational paradigms required to establish them. %

\subsection{Attachment}
\label{sec:attachment}

Attachment theory \cite{bowlby1969} is, overwhelmingly, the theory most-often invoked in the study of human--robot relationships \cite{mitchell2025}. Bowlby's formulation was undertaken by joining ethology to control-systems principles, conceptualising the attachment system as a homeostatic controller. The system monitors accessibility of a protective figure, and activates proximity-seeking behaviours when threat\footnote{Those conditions are broader than the word ``threat'' suggests. Bowlby's list includes fatigue, illness, pain, hunger, alarm, unfamiliarity, and the figure's own unavailability or non-responsiveness. The list is non-exhaustive, and we use ``threat-gated'' henceforth as shorthand for those conditions, since a distressed user (or a baby) at three in the morning is fatigued and alone.}  or separation drives that appraisal below a set-point \cite{bowlby1969,ainsworth1978,granqvist2021}. While Bowlby originally framed the set-goal as physical proximity—and later as appraised \emph{availability}—the standard contemporary reading of the set-point as \textit{felt security} reflects the organisational revision introduced by \cite{sroufewaters1977}. The system is characterised by what it is \emph{for}. Its set-goal is security, and its outputs are proximity-seeking (of the attachment figure). This also manifests as; protest at separation, use of the attachment figure as a \emph{safe haven} under threat, and use of that figure as a \emph{secure base} from which to explore when threat is absent \cite{waters2000,cassidy2016}. Attachment, as it was canonically defined, is therefore not a synonym for any (seemingly) close relationship. It is a specific (and demanding) \emph{type} of relationship, though available in principle to any dyad that meets its criteria. Ainsworth \cite{ainsworth1989} drew this distinction more sharply, separating \textit{attachment bonds} from \textit{affectional bonds} in general. For an \textit{affectional bond}, the criteria was as follows:

\begin{enumerate}
    \item the bond \textbf{persists} rather than being transitory
    \item  it is \textbf{specific} to a particular figure who is not interchangeable with another
    \item  it is emotionally \textbf{significant}
    \item the individual wishes to maintain \textbf{proximity} to or contact with that figure
    \item involuntary separation causes \textbf{distress}.
\end{enumerate}

She suggested that an affectional bond becomes attachment \textbf{if and only if} the partner is sought as a source of security and comfort, and serves as a base from which the individual can move off to explore (the ``security'' criterion). Therefore, a selective, enduring, rewarding, proximity-maintaining relationship that lacks the safe-haven and secure-base functions would not qualify as an attachment under this definition \cite{weiss2006,sheldon1989}: this holds, no matter how seemingly-close this social bond appears to be. It is criterion 6 (security provision) that any HRI study claiming an ``attachment'' must meet, and also the criterion carrying the evidential and ethical weight for research looking to understand such a relationship.

That affiliation and security provision disconnect has been observed empirically in biological dyads. In adult humans, security-provision functions cluster together and separate from the intimacy and disclosure that distribute across close relationships more broadly \cite{sheldon1989}. In fact, adolescents' own descriptions distinguish the figures they turn to for security from those with whom they are simply affiliated \cite{martin2017}. Comparative work makes the same separation on physiological grounds \cite{mendoza1997,witczak2023}, as does developmental theory. This latter perspective distinguishes the selection of a discriminated figure---which human bond formation shares with animal bonding---from the criterial secure-base achievement, which it does not \cite{freeman2024}.

In the context of relationships between human and artificial systems, this disconnect means that attachment cannot be inferred from (perceived or actual) high user engagement or interaction time alone. Claiming that attachment is present requires empirical evidence of threat-gated security functions, specifically by demonstrating that the artificial partner serves as a safe haven under distress, and a secure base for exploration. Where such relationships lack (the evidence for) these security functions, assuming an attachment risks mischaracterising the underlying behavioural system. However, many relationships that fall short of ``attachment'' are not necessarily devoid of relational substance. Some may still exert distinct, partner-specific effects on an individual's emotional and physiological state, which may instead be characterised as an affective social bond.

\subsection{Affective Social Bonding}
\label{sec:bonding}

Where attachment is organised around felt security, a social bond is organised around affective regulation. Explicitly, it is an arrangement in which a \textit{particular partner regulates the organism's affective and physiological state}. As introduced in Section~\ref{sec:intro}, we use \textbf{affective social bond} to designate this definition, and \emph{social bond} as shorthand for it here. It extends the construct from Ainsworth's \emph{affectional bond}, a broader term defined by persistence, specificity, emotional significance and separation distress, with reference to what the partner regulates (Section~\ref{sec:attachment}). We also distinguish this from the flexible use of ``bond'' in HRI which has applied to ties that show changes in social interaction or attitudes towards a robot. It also marks where we converge with \citet{konijn2025}, for whom bonding to a social robot is a matter of feeling emotionally or affectively connected. 

In the perinatal and nursing literatures, \textit{bonding} indicates a parent's affective self-reported orientation toward an infant, directed from parent to child \cite{ettenberger2021,kinsey2013concept}. In the ethological and neuroendocrine literatures, it suggests a selective, affiliative preference for a particular individual, sustained by identifiable affective and physiological mechanisms. These include stress buffering \cite{kikusui2006,carter1998}, behavioural synchrony \cite{feldman2012,young2001}, and thermal and cardiac co-regulation \cite{fotopoulou2022}. Here, we use the second definition, providing a basis for the procedure that follows in Section \ref{sec:operationalising}. It provides measurement operations that does not (only) run through participant self-report, thus allowing for cleaner distinction between the two systems in question. 

In that sense, an affective social bond functions as regulation of affective state through a particular partner. This brings about two requirements. First, that the arrangement pertains to \emph{this} partner, and secondly, that this particular partner measurably does something to the organism's body. We refer to these requirements as ``\textbf{selectivity}'' and ``\textbf{regulation}'' respectively, and discuss each in more detail below.

\subsubsection{Criterion I: Selectivity}
\label{sec:selectivity}

Selectivity is the minimum condition for a bond, held in common across every tradition that define the phenomenon; in fact, these traditions converge on it independently. It is Ainsworth's second criterion, which requires a figure ``not interchangeable with another''. It was evident in the first longitudinal study of attachment formation, in that the figure an infant selected was not reliably its primary care-giver (i.e. the one performing most routine care) \cite{schaffer1964}. That dissociation, between the partner an organism is \textit{oriented to} and who it has most \textit{exposure to}, constitutes the core of the selectivity criterion. Clearly, this makes exposure unusable as a proxy for selectivity. In turn, this bears on the human--robot case directly, since interaction time and exposure dominate current studies \ref{sec:hribond}. 

The test that operationalises this criterion comes from a range of biological dyads. These require preference for a familiar partner over a stranger, and score it as a choice against a matched alternative \cite{williams1992ppt,carter1998,beery2021ppt}. The same contrast separates a bond from general affiliation in the human peer and primate literatures \cite{mendoza1997,samuni2021,wittig2014}. This is also distinguished in adult humans, where security-provision functions cluster together and individuate intimacy and disclosure distributed across close relationships more broadly \cite{sheldon1989,martin2017}. Partner preference is supported by oxytocin, vasopressin and dopamine mechanisms coupling the reward value of a familiar partner to a mesolimbic circuit \cite{winslow1993,insel1995ota,ross2009otr,walum2018}. Selectivity is also dissociable from affiliation. Accumbal dopamine release is itself partner-specific, i.e. greater for partner-directed seeking/interaction than for identical behaviours toward a novel partner \cite{pierce2024selective}.\footnote{We cite the mechanism here as evidence that this criterion is implemented in biological dyads, and not as a recommendation about which peptide to measure. We offer it also as justification for the observable and practical proxies that we propose in Table \ref{tab:reg} }.

Just as `attachment'' is not the default description of a strong tie, sociability is not the same as selectivity. Watching an animal approach a conspecific does not imply that it is bonded to it, no matter how social it appears to be. Similarly, a robotic system that any user finds trustworthy or pleasant to interact with should not be taken as evidence of selectivity. Meeting the selectivity criterion requires, at minimum, an alternative agent that an individual \textit{could} have preferred instead, not simply changing interaction dynamics to a particular (single) agent.

\subsubsection{Criterion II: Regulation}
\label{sec:regulation}

The second criterion characterises the bond by what the partner does---physiologically---to the other's body. We consider this the ``affective'' mechanism of affective social bonding. To say that a person feels connected to an agent \cite{konijn2025} is to describe a particular affective state. Affective states are implemented physiologically, and where an affective state depends on the presence of one particular other, that dependence is measurable. This, we propose, is the criterion that can most sharply separates the construct of affective social bonding from an attachment. \textit{Regulation} is what the affect in the ``affective social bond'' consists of at the level that the body implements it.

In this view, a bonded partner is an external regulator of the organism's internal state. Physiological set-points that an individual cannot sustain alone are held, in part, by the reliable presence of that particular other, so that maintaining a bond is maintaining a shared regulatory system. As with selectivity, oxytocinergic signalling is the mechanism most often implicated in this arrangement (see footnote 2). For instance, by potentially acting with dopamine on a mesolimbic reward pathway to consolidate the reward value of a \textit{particular} partner \cite{liu2003otda,ross2009otr,young2001,walum2018}, and (separately) acting on both the hypothalamic--pituitary--adrenal (HPA) axis and on amygdalar reactivity, dampening stress responses \cite{heinrichs2003,kikusui2006,hostinar2014}. This idea originated with Hofer 's work on rodents\cite{hofer1973,hofer1994}, showing that a mother regulates several components of her infant's physiology (e.g. cardiac rate, thermogenesis, and metabolic activity). These all occur through separable sensory channels hidden inside a dyadic interaction. In fact, the absence or loss of the mother is the withdrawal of several independent regulators at once. This gave rise to generalised accounts of bonding as \emph{co-regulation}, in which bonded partners influence each others' physiology through various (separable) mechanisms to maintain one another's homeostasis. Contemporary accounts give this control-theoretic form, with social baseline \cite{cohen2004} and social homeostasis \cite{matthews2019} theories treating proximity to a trusted other as an individual's \textit{expected} condition. The brain budgets its own regulatory effort against this condition, and its absence as a metabolic cost \cite{beckes2011,coan2006,matthews2019,schulkin2011,sterling1988,mcewen1998}. 

Being bonded, on this reading, is a state in which the body's regulatory settings have been adjusted on assumption that the (preferred) social partner will be present. Unlike attachment, which is organised around the single set-goal of felt security (Section~\ref{sec:attachment}), bonds are individuated by the mechanisms that implement them (Section~\ref{sec:regulation}).  Separability of these regulators, both \textit{within} the same social bond and \textit{among} different social bonds, offers numerous physiological and biobehavioural regulators that can be measured independently in an experimental setting (see Table \ref{tab:reg}).

That said, listing all potential physiological regulators that can be present in a bonded dyad is beyond the scope of this paper (however, see \cite{feldman2017,carter2009} for a more thorough discussion).  Below, and in Table~\ref{tab:reg}, we offer a working inventory that we propose would be well-suitetd for HRI experiments. The list is non-exhaustive, and one which we have previously discussed in part \cite{khan2024}, grounded both in their evidentiary strength and practicalities of implementation for HRI practitioners.

\textbf{Social buffering} is the attenuation of an organism's physiological stress response in the presence of a particular partner. It is implemented by dampening of the HPA axis and suppression of corticotropin-releasing factor \cite{hostinar2014}, with oxytocin acting as a modulator of that dampening \cite{heinrichs2003}. Oxytocinergic signalling is implicated in buffering across species \cite{carter1998,kikusui2006}, as well as cross-species \cite{beetz2012dogbuffer}. It is typically expressed as reduced cortisol reactivity and faster recovery under a standardised stressor and, at finer temporal resolution, as lowered electrodermal activity and elevated heart-rate variability while a bonded partner is present. These can be captured via wearable electrodermal and cardiac sensing hardware, neither of which are new to HRI experimental paradigms \cite{gupta2024human,perugia2017electrodermal}. We propose that this accessibility and (relative) low-cost can be a promising avenue to pursue for HRI researchers.

\textbf{\emph{\textit{Synchrony}}} is the temporal coupling of two partners' movement, gaze, posture, microgestures (in the case of ``behavioural synchrony'') and autonomic rhythms (in the case of ``physiological synchrony''). It is implemented by entrainment between coupled biological oscillators \cite{feldman2007synchrony, oullier2008coordination}, and it is the channel on which the bonding literature has leaned hardest. Further, biobehavioural synchrony is also proposed as a process through which affiliative bonds are formed \cite{feldman2012,feldman2017}. It is expressed as cross-correlation of movement, speech and cardiac phase, and analysed via windowed cross-correlation against surrogate controls \cite{ramseyer2011mea,moulder2018surrogate,behrens2020wcc}, with open implementations available \cite{hudson2021multisyncpy}. Synchrony is also a cost-effective option to record against an artificial partner. Video and motion tracking together with the agent's own data log (e.g. movement or event flags, motor positions, trajectories) are sufficient, and HRI has adopted this measure, at least in part \cite{stoeva2024hrisynchrony,lorenz2011synchrony}. 

\emph{\textit{\textbf{Separation distress}}} is the affective response to a bonded partner ceasing to be available. It is implemented in the body by withdrawal of endogenous opioid signalling together with corticotropin-releasing factor activation \cite{panksepp1978,loseth2024}. It is expressed as negative affect, searching, protest and grief on removal. Much like synchrony, its behavioural proxies are sufficient to index changes in the physiological substrate. These responses have also been documented for artificial systems \cite{defreitas2024,banks2024}.

\emph{\textbf{Pain and threat modulation}} is the attenuation of an organism's subjective and physiological response to noxious stimuli in the presence of a partner. It is implemented by endogenous $\mu$-opioid signalling \cite{machin2011,tchalova2019}: in context, a partner's handholding lowers threat-related neural responses where a stranger's does not, and slow affective touch reduces both reported pain and its neural signature \cite{coan2017handholding,vonmohr2018pain,pan2024touch}. It can be expressed as higher pain thresholds, lower subjective intensity ratings, and dampened autonomic responses (such as skin conductance) during a standardised noxious event. Data for this channel has already been collected in HRI using thermal heat pain paradigms \cite{geva2020}, electrocutaneous threat anticipation \cite{yim2022}, and targeted affective touch \cite{nakae2025}. 

\emph{\textbf{Selective aggression}}, or mate-guarding, is the active defence of a bonded partner against novel intruders or rivals. It is implemented in mammalian models (e.g. prairie voles) via vasopressinergic signalling, which supports both partner preference and aggressive rejection of strangers \cite{winslow1993}. The ``intruder paradigm'' used in animal models may not be ethically run on humans, though its functional analogues of interpersonal jealousy, mate-guarding, and distress over breaches of exclusivity \cite{pfeiffer1989jealousy}, may be adapted for HRI scenarios (see Section \ref{sec:q3}).

Finally, \textbf{thermal and cardiac co-regulation} is the holding of autonomic set-points by a partner's physical presence. For example, skin-to-skin contact stabilises a neonate's heart rate and temperature \cite{fotopoulou2022}. Unlike social buffering, which is read against response to an applied stressor, co-regulation is visible at rest and appears as set-point drift once the partner is withdrawn. This co-regulation is delivered through thermal, tacticle, and vestibular channels, directly by the partner themselves. This ties them to contact, subsequently, to embodiment and thereaafter to a morphology that affords it (Section~\ref{sec:embodiment}). For that reason we do not consider these two channels feasible to operationalise across the current landscape of available robots.%

\subsubsection{A Note on ``Regulation'' as a Criterion}
\label{sec:regulation_note}

Although the comparative literature sometimes individuates a social bond behaviourally (i.e. by partner preference) \cite{carter1998,walum2018,beery2021ppt}, we emphasise the regulation criterion for reasons specific to artificial systems. That is, selectivity is easy to fake.  A system or robot that retains a per-user memory, addresses the user by name and recalls their history will produce individuation, apparent preference and the entire history and vocabulary of a relationship. In short, an entire persona can be replicated. The same is true for (self-reported) affect. An agent designed to \textit{elicit} felt connectedness may elicit it, and study designs that rely on self-reports would report as such. What a transferable persona cannot manufacture is a change in the user's physiology that depends on \emph{this specific agent}. Physiological regulation is the criterion that distinguishes the presence of a bond with a specific social agent from something elicited from a design feature; thus, we consider it a hard criterion for measuring social bonding.

\subsection{Cross-Species Validity and Transferability to HRI}
\label{sec:transfer}

HRI practitioners may reasonably ask why criteria established in prairie voles, titi monkeys and human infants should be expected to say anything about a person and a machine. Our response is that these diagnostic paradigms have already successfully crossed non-human species boundaries. Rather than remaining restricted to intra-species human relationships, attachment and bonding frameworks have been extensively validated in human--animal dyads. For instance, proximity maintenance toward companion animals is often expressed as strongly as toward fathers or siblings, with secure-base and safe-haven dynamics emerging as primary structural features of the tie \cite{kurdek2008pet}. Under rigorous testing, non-human partners reliably fulfill the secure base and safe haven functions \cite{zilchamano2012}, and physiological regulation (here, stress buffering) in both children \cite{beetz2012dogbuffer} and adults \cite{gandenberger2024presence}.

The core premise of our framework is that these experimental procedures are inherently transferable. Having already crossed the biological species boundary into human--pet relationships, the human--robot dyad represents the logical next application. Existing evidence offers no theoretical reason to assume that these biobehavioural regulatory mechanisms would operate differently in kind for an artificial agent. The gap is simply one of empirical rigour. Extending attachment criteria to non-biological targets has explicit precedent in the attachment-to-God literature. \citet{granqvist2010religion} demonstrate that believers' relationships with God ``meet the definitional criteria for attachment relationships''—despite a target that is non-corporeal, non-reciprocating, and physically absent. Whatever one concludes about the substantive claim, the methodological precedent matters here, since the extension was argued criterion by criterion, rather than by analogy, and the criteria were held fixed while the target changed. That is the form the artificial case has to take as well. Whether these procedures yield interpretable results in HRI is a testable, empirical question of experimental methodology—entirely distinct from the philosophical debate of whether a machine can be bonded to in an ontological sense.

\section{How Human--Robot Interaction Has Used ``Attachment'' and ``Bonding''}
\label{sec:hriuse}
\FloatBarrier
We now turn to HRI's treatment of the terminology defined in previous sections, focusing both on how they have been discussed/used within the field, and how they disconnect from the definitions/criteria set out above. The overview below is not exhaustive, but we are nevertheless in agreement with several authors \cite{mitchell2025,law2022,konijn2025} that HRI has used both terms loosely, resulting in a free-for-all of definitions, methods, and conclusions made about human--robot relationships. %

\subsection{``Attachment'' in HRI}
\label{sec:hriattach}

\citet{mitchell2025} define human--robot attachment as the ``emotional bond between humans and robots facilitated by prolonged interaction, a unique robot identity, and effective communication''. Comparing this definition with Ainsworth's six criteria in Section \ref{sec:attachment}, prolonged interaction can be mapped onto persistence (criterion 1), and unique identity onto specificity (criterion 2). Their definition of effective communication maps onto none of the criteria, and the authors' definition also fails to address emotional significance (criterion 3), proximity maintenance (criterion 4) separation distress (criterion 5), or that security be sought (criterion 6, and the defining criterion for ``attachment''). Whilst the authors therefore address two of the six original criteria in their definition, what they specify would actually categorise something weaker than an affectional bond in Ainsworth's construction. Because ``there are no current definite definitions of attachment in HRI research'' \cite{mitchell2025}, their review's search string pooled five terms---human--robot \emph{attachment}, \emph{bond}, \emph{rapport}, \emph{friendship} and \emph{companionship}---into a single corpus. Therefore, whatever the reviewed studies have in common, they cannot be considered a single, unified construct, because five separate ones were pooled at the point of retrieval. That is not a criticism of the review, which was constrained by the literature available. Rather, it is a measure of how little the field's current vocabulary discriminates. \citet{banks2026mc} since reviewed the fifth of those terms individually, and their finding (that there were over fifty distinct measured variables across seventy one studies) further emphasises how fragmented the terminology and methodologies for studying attachment are.

A second definition, traced by \cite{law2022} through much of the field, comes from outside psychology, using \cite{norman2004}'s definititon. Here, ``attachment'' to a device is the cumulative sum of the emotional episodes a user has had with it. However, this definition does not appear to directly map onto any of Ainsworth's criteria. Rather, it is a quantity of subjective feeling, collapsed into a single value.

Conceptual confusion in HRI stems largely from overloading the term ``attachment,'' which is frequently conflated across three distinct levels. There are a \emph{pre-existing trait} of the human \cite{dziergwa2018cohabitation, gillath2021attachment,yang2025}, an \emph{evaluative attitude} toward an artefact \cite{koverola2022gators,naneva2020review,vanmaris2020ethical,collins2013assistive}, and a \emph{relational status} \cite{van2020child}. This ambiguity is well-documented in HRI scholarship \cite{kahn2006aibo,melson2009robotic}. While relational-artifact literature posits that interactive agents occupy a distinct ontological category unsuited to human interpersonal terms \cite{turkle2006relational}, our framework translates this theoretical position into an empirical test (Section~\ref{sec:operationalising}). Whether a human--robot tie represents a novel relational category depends entirely on which biobehavioural criteria it satisfies under experimental observatio. Thus, it requires a standardised procedure, capable of identifying interaction profiles that fit neither classical human attachment nor simple object use.

This disassociation is reflected in current measurement instruments. Two contemporaneous self-report inventories both bear the name \emph{AI Attachment Scale}, despite sharing no factor structure \cite{cheng2026,kasturiratna2025}. Furthermore, while the secure-base subscale in \cite{cheng2026} targets the relevant criterion, its reliance on Likert items measures self-reported \emph{perception}, rather than functional biobehavioural operation. In a similar vein, \cite{chen2026} state the strict attachment criteria, acknowledge that no human--robot dyad currently meets them, yet paradoxically conclude the tie is "attachment-like" and revert to a cumulative scalar model. %

Returning to the review of \citet{mitchell2025}, all thirty included studies used subjective measures; nine drew on established psychological instruments. Among them a romantic-love inventory, a customer-service rapport subscale and a product-attitude scale, used to measure the same named construct. A further eight used original or unvalidated measures, and three adapted human--pet scales. Its verdict is that these instruments ``do not measure attachment directly but instead assess factors associated with attachment'', and ``often fail to incorporate the psychological frameworks of attachment such as understanding whether the robot is seen as a secure base or a safe haven, and assessing proximity-seeking behaviour and separation distress''. What these instruments record, then, is an attitude toward a \emph{class} of artefact rather than a relationship to a particular agent. This represents the difference between asking someone how they feel about cats in general, and asking them about their own lifelong pet.

\subsection{``Bonding'' in HRI}
\label{sec:hribond}

To date, measures in HRI related to ``bonding'' establish neither requirement set out in Section \ref{sec:bonding}.  The substitution of trust for bonding is one of the most common failures \cite{khan2026}. In the widely-cited paper of \cite{belpaeme2012}, which carries the phrase ``building social bonds'' in its title, the ``bond'' was a single forced-choice closeness item. The participants who placed the robot at ``acquaintance'' or ``stranger'' were mainly those who had experienced a malfunction; a co-occurrence equally consistent with the measure tracking reliability. This dilution is also seen across long timescales. \cite{tanaka2007socialization} immersed a social robot in a toddler classroom for over five months, observed that children came to treat it as a peer, and concluded that the technology was close to achieving autonomous \emph{bonding} with toddlers. Despite a careful observation, what was measured in their study was interaction quality over time, from which bonding was inferred. The same inference is taken as the premise of ongoing child--robot work, framing self-disclosure and warmth as bonding \cite{neerincx2021childbonding}. The substitution appears on extremely short timescales, in a study that operationalised the bond as a single item administered after two one-way robot speeches of under three minutes each, in a design whose reported outcome is the effect of perceived similarity on trust \cite{salekshahrezaie2021}. More recent instruments name a bond and then measure something adjacent to it. \citet{diazboladeras2023} defines bonding as the emotional relatedness that unfolds over time'' and decomposes it into engagement and social rapport; \citet{yankouskaya2025} name the relational subscale of {LLM-D12} ``parasocial bonding'', and define the construct as forming parasocial bonds that reflect perceived companionship, emotional closeness, and social fulfilment, while its items ask about self-disclosure, companionship and loneliness. \citet{barakova2018}'' operationalises bonding as twelve items adapted from an immersion questionnaire, split into game engagement and empathy with the partner. None of these is a bad instrument for what it measures, per se. What none of them establishes is that the tie is to a specific partner, or that the specific partner supplies any affective regulatory effect.

Current evidence on human--robot bonding spans two extremes, but does not answer the question these criteria pose. At one end is superficial anthropomorphism: owners of non-social robotic vacuum cleaners name them, assign them genders, and alter their homes to accommodate them \cite{sung2007roomba}. Because users apply relational vocabulary this readily to a cleaning appliance, the presence of such language in social robotics cannot, on its own, be evidence that a bond has formed. It records how people describe a relationship, not whether the relationship meets specific criteria. At the other end is the physiological evidence from companion robotics, which speaks to regulation but not selectivity. A care-house deployment of two therapeutic seal robots reported increased social interaction among residents, alongside improved urinary stress-hormone profiles, over the first month \cite{wada2007paro}. This remains among the field's best evidence on the \emph{regulation} requirement, though without a control condition that would separate that specific robot from the attention to a novel robot in general. A later cluster-randomised trial then provided that control, comparing the same robot against the same robot with its robotic features disabled, and against usual care \cite{moyle2017paro}. However, its outcomes were behavioural and observational---engagement, affect, agitation and quality of life---so it still lacks evidence for the \textit{regulation} criterion set out above. The two therefore trade off against each other: the study with the physiological outcome lacks the control, and the study with the control lacks the physiological outcome. Neither addresses \emph{selectivity}, since outcomes in both are reported at group level for residents interacting with a shared artefact. This leaves the question as to whether any resident's state depends on \emph{that particular} unit rather than a robot of that kind. That regulation can be reported, while selectivity goes untested, is itself informative, because it shows the two criteria to be dissociable.

Much of this evidence also rests on brief HRI encounters. Of the thirty studies in the \citet{mitchell2025} review, sixteen ran for less than a day and only five for two months or longer. Surveys of long-term interaction find the same pattern, with novelty accounting for much of early engagement and interest declining beyond a few weeks. Recent reviews show how small the corpus of extended deployments remains \cite{matheus2025longterm,robinson2019rct}. Simply put, an affective social bond--a phenomenon defined by ongoing regulation and enduring selectivity---cannot be established in a single day, yet more than half of the field's evidence on social relationships was collected in this way.

Finally, a distinct tradition measures bonding as behavioural coordination. Drawing on interpersonal psychology, where movement and postural synchrony are measurable indices of a bond \cite{oullier2008coordination,fujiwara2020rhythmic}, HRI has adopted this paradigm using time-series analysis and explicit models \cite{lorenz2011synchrony,prepin2007synchrony}. While behavioural synchrony is indeed a significant component of social bonding, which we also adopt in our proposed framework (Table~\ref{tab:reg}), behavioural entrainment frequently occurs between strangers \cite{jouaiti2024synchrony} and misses the requirement of \textit{selectivity}. These works do not establish that synchrony is tied to a particular agent, and without testing for it, claims of a social bond in such dyads cannot be made.

Nonetheless, some uncontrolled \emph{observations} in and around HRI nonetheless satisfy some of Ainsworth's criteria, and do so in populations with no incentive to anthropomorphise. Explosive-ordnance-disposal operators name their units, ascribe them personalities, express distress at their destruction, and in some cases decline a functionally identical replacement in favour of having their own unit repaired \cite{carpenter2016eod}. Discontinuation of companion applications produces grief on the scale of a human loss, and personality-altering updates produce reports of a specific individual having been lost \cite{defreitas2024,banks2024,zhang2026fragility}. These are evidence of affectional-bond components, and all criteria being satisfied in a population with no incentive to anthropomorphise or treat their units as social agents.

\subsection{Against the ``magnitude'' assessment of social relationships} %

\label{sec:magnitude}
\label{sec:compound}

Previous efforts in HRI and adjacent fields have frequently sought to collapse distinct relational constructs onto a single scale, treating attachment and bonding as the same \textit{type} of relationship expressed at varying magnitudes \cite{rabb2022,griffin2025,szondy2024attachment,eum2026}. Others model attachment as a cumulative sum of affective experiences with an artefact \cite{norman2004,chen2026}—a definition imported from consumer product design and, as \citet{law2022} document, adopted across HRI with little critical scrutiny. At the extreme limit, attachment is computed as a scalar quotient derived directly from interaction parameters such as intimacy or engagement \cite{samani2010}. The appeal of this scalar reduction is obvious: a single metric can be tracked across longitudinal deployments, and compared across experimental conditions. It can also be entered cleanly into statistical models, and dynamically integrated into personalised robot architectures.

However, these magnitude-based approaches rest on the flawed assumption that underlying components of a social tie co-vary (i.e. that an increased score on one dimension represents a net increase in overall relationship "strength"), or that crossing a quantitative threshold transforms the nature of the tie. Cross-species neurobiological and behavioural evidence directly contradicts this, demonstrating that security provision, affective regulation, and social preference rely on distinct, dissociable mechanisms \cite{kendrick1997, lee2019selectivity,sailer2022dissociation, hofer1994, hostinar2015adolescence,gunnar2015,hostinar2014,doom2016}. Consequently, while computationally convenient, reducing a relationship to a single scalar value risks misrepresenting both the qualitative type and specific biobehavioural reality of the human--robot tie.

The strongest counterargument in defense of scalar measurement is articulated by \citet{mitchell2025}. Their review accurately states Ainsworth's criteria, noting that the "strict definition helps ensure that only the deepest and most supportive relationships are classified as attachments." However, they argue that applying this strict definition in HRI "may be limiting and not fully capture the range of human-robot relationships," ultimately endorsing the continuous spectrum proposed by \citet{rabb2022} to prevent weaker ties from being excluded from empirical study. We agree entirely with their starting premise. Weak ties constitute the majority of human--robot interactions, and a framework capable only of returning a binary "not an attachment" verdict would be of little use to the field. Where we disagree is the remedy. When a scientific construct proves too narrow for a broader set of phenomena, the appropriate solution is to introduce distinct, dedicated constructs—not to stretch the existing one until it loses all precision. The outcome of stretching the construct is demonstrated within the review itself. Widening a single term did not illuminate the diversity of human--robot ties; it rendered the entire spectrum methodologically intractable. We argue that a far more rigorous alternative is to define these relational types as discrete, testable classes—each governed by its own evidential criteria—and to move away from collapsing distinct biobehavioural systems into a single, unidimensional magnitude.

\section{A Procedure for Classifying Human--Artificial Ties}
\label{sec:operationalising}
\label{sec:classify}
\FloatBarrier

Building on the criteria established in Section~\ref{sec:home}, we propose a procedure that allows for specific classification of a particular human--robot relationship. The procedure is presented as a set of four questions, which can be asked of a given human--robot relationship, along with explicit evidential requirements necessary to answer each. Throughout, we use \emph{relationship} as an unclassified, baseline term designating whatever tie exists between a specific person and a specific artificial agent before empirical evaluation. \emph{Attachment} and \emph{affective social bond} are subsequently treated as specific verdicts that require fulfilling their respective empirical burdens.

These questions exhaust the criteria discussed above. Ainsworth's five-plus-one criteria are distributed across Questions~1, 2, and 4, social bonding's \textit{selectivity} requirement is included in Question 2, and the \textit{regulation} requirement is included in Question~3. By including the means to answer these questions in experimental design and/or data analysis, we hope that the field of HRI can begin to find some standardisation and consistency when making claims about human--robot relationship.

The questions are as follows. For a particular human--robot dyad, ask:

\begin{enumerate}[start=1,leftmargin=1.6em,itemsep=0.1em,topsep=0.25em]
\item \textbf{Is there a tie to classify?} Does the dyad show the \textit{observable} marks of an affectional tie---persistence, emotional significance, voluntary proximity maintenance, and distress at involuntary separation? \emph{(Ainsworth criteria 1, 3, 4 and 5)}
\item \textbf{Is the relationship selective?} Does the human orient to \emph{this} agent in preference to a matched alternative? \emph{(Bonding Criterion I; Ainsworth criterion 2)}
\item \textbf{What does the agent regulate?} Which of the physiological or affective regulators does this specific agent provide? (Table~\ref{tab:reg})  \emph{(Bonding Criterion II)}
\item \textbf{Are the security functions activated?} Is this specific agent sought under an activating condition, and is exploration organised with respect to its availability? \emph{(Ainsworth criterion 6)}
\end{enumerate}

In practice, this reduces to three rules:

\begin{enumerate}[start=1,leftmargin=1.6em,itemsep=0.1em,topsep=0.25em]
    \item A study claiming any social relationship in a human--robot dyad must have answered positively to Question 1.
    \item A study claiming an \textbf{attachment} must have answered Questions~1, 2, and 4 positively. 
    \item A study claiming an \textbf{affective social bond} must have answered Questions~1, 2, and 3. 
\end{enumerate}

The answers to these questions return one of six statuses, described in Table~\ref{tab:classify}. The two of interest in the current paper---\textit{attachment} and \textit{affective social bonding}---are included. Where partial or incomplete answers to these questions are obtained, we offer several related classifications, in line with Ainsworth's criteria. Specifically, \textit{no relationship to classify, non-specific regulation, affectional tie, affiliative preference}. We propose that these relationship type should be treated as separate categories. This categorisation makes no claims regarding the magnitude or strength of a particular relationship, and we do not suggest comparative statuses, such as being ``more'' or ``less'' bonded relative to an alternative.

Questions 1, 2, and 4 can return the answers of \emph{met}, \emph{not met}, and \emph{not tested}. Specifically tracking \textit{not tested} is an important addition. A study that, for example, fails the (attachment) security test, and one that does not run the test in the first place, should not get treated in the same way in the classification of a relationship. We consider that being able to state exactly what was/was not established as part of a study (with respect to a specific procedure) is an important move towards standardised reporting. We provide the example below of a possible (non-prescriptive) reporting style that can be used: %

\begin{quote}\small
\emph{Q1 (is there a tie to classify?)} Met. Seven-month deployment; contact logged behaviourally; distress recorded at a scheduled outage; significance self-reported.\\[0.15em]
\emph{Q2 (is it selective?)} Yes.\\[0.15em]
\emph{Q3 (what does the agent regulate?)} Stress buffering supplied, agent-present against agent-absent; no matched alternative; other channels not tested.\\[0.15em]
\emph{Q4 (do the security functions obtain?)} Not tested.\\[0.15em]
\emph{Status returned:} \textbf{Affective social bond.}
\end{quote}

Though enumerated, these questions do not need to be answered in order, with one exception. Question~1 establishes whether any tie exists at all, so a negative answer to it---no observable persistence, emotional significance, proximity maintenance, or distress at involuntary separation---means the relationship cannot be an attachment or a social bond, regardless of the answers to remaining questions. It therefore acts as an entry condition in classification of a social relationship. Unlike the other questions, however, it can be answered entirely by observation and record, rather than by experimental manipulation (see \cite{carpenter2016eod,sung2007roomba,defreitas2024}). Additionally, Question~3 is answerable of any dyad, and a positive answer to it alone is possible: a responsive artefact can alter physiological state without that state depending on the particular unit \cite{wada2007paro}. Question~4 entails Question~2 by design, because the safe-haven and secure-base functions are defined with respect to a specific figure. Nothing else follows in the other direction. That is, selectivity does not stand for regulation (and vice versa), and regulation is not evidence of security (and vice versa).
 
Sections~\ref{sec:q1} to~\ref{sec:q4} take each question in turn, describing what it establishes, and what a study must do to assess it. For each question, we offer representative study designs drawn from paradigms already validated in the human--human, human--animal and animal--animal literatures and, where possible, an entry-level experimental design that can be run more easily.
 
\subsection{Question 1. Is there a relationship to classify?}
\label{sec:q1}
 
Question~1 establishes that the dyad shows the observable marks of an affectional tie, carrying four of Ainsworth's five criteria. These criteria can be met through observation alone, and do not require direct experimental manipulation, allowing it to be answerable of any human--robot dyad with relatively low effort:
 
\begin{enumerate}[label=(\roman*),leftmargin=1.9em,itemsep=0.05em,topsep=0.2em]
\item \emph{Persistence.} The relationship is observed beyond the novelty window. HRI's own long-term data put that at a few weeks at minimum, since novelty accounts for much of early engagement and interest declines beyond it \cite{leite2013longterm,degraaf2016homes}.
\item \emph{Proximity maintenance.} Contact with the agent is sought voluntarily, recorded behaviourally, and not inferred from continued use of a device the person depends on for something else (e.g. a smartphone).
\item \emph{Separation distress.} Distress is recorded at an \textit{actual} (rather than hypothetical) interruption, either planned or naturally occurring.
\item \emph{Emotional significance.} The person (self-)reports the relationship with the agent as mattering to them.
\end{enumerate}

\subsubsection{Potential Study Designs}
\label{subsec:q1_designs}
Studies that answer this question already exist across much of the human--artificial agent literature. Discontinuations and personality-altering updates of companion applications record separation distress at scale---uncontrolled, but at a magnitude that is hard to attribute to novelty \citep{defreitas2024,banks2024,zhang2026fragility}. Multi-month deployments establish persistence \cite{tanaka2007socialization,wada2007paro,scassellati2018}. Proximity maintenance and emotional significance are recorded together in the ethnographic literature on domestic and field robots, where owners name their units, adapt their homes around them, and describe the relationship as mattering \cite{sung2007roomba,carpenter2016eod}. With these examples in mind, a prospective study needs a deployment long enough to clear the novelty window \cite{leite2013longterm} (e.g. months-long deployment \cite{scassellati2018}, a behavioural record of contact (e.g. ethnography \cite{jacobs2020evaluating}) or digital event logs, interruption of contact (withdrawal of the agent \cite{yamazaki2023long}, a planned maintenance window, or a naturally occurring outage) to induce separation, and one self-report item on whether the relationship matters, as a means to assess emotional significance \cite{rosenthal2013experimental}.

Following from the discussion by \cite{weiss2006,hazanzeifman1994}, all four of these criteria must be met of any relationship that will be defined as either an attachment or a social bond. In absence of any one of these criteria being met, the relationship may still fall under a different classification, but such classifications fall beyond the scope of this paper. 
 
 \subsection{Question 2. Is the relationship selective?}
\label{sec:q2}
 
Question~2 establishes that the person's orientation is to \emph{this} specific agent, and not only to agents of its kind.  A human partner that appears to show increasing sociability, changes in attitude or improved engagement towards a single robotic agent would not, on that evidence alone, suffice as evidence for selectivity; the same response might be produced by any agent of that type. A study designed to answer this question requires:

\begin{enumerate}[label=(\roman*),leftmargin=1.9em,itemsep=0.05em,topsep=0.2em]
\item an \textit{alternative agent} that the person \textit{could }have preferred, matched on capability, availability and/or appearance;
\item that this alternative be another instance of the \emph{same kind}---ideally a physically identical unit---so that the person's \emph{interaction history} is the only systematic difference between the two.
\item  the changes in interaction are measured behaviourally, e.g. contact time, interaction time, or a discrete choice
\end{enumerate}

\subsubsection{Potential Study Designs}
\label{subsec:q2_designs}
Several design options in the animal literature can be adapted for human--robot scenarios. Firstly, an \emph{adapted partner-preference test.} The animal paradigm places the subject between a familiar partner and an unfamiliar conspecific with free access to both and scores side-by-side contact time \cite{williams1992ppt,carter1998,beery2021ppt,mendoza1997,samuni2021,wittig2014}. In HRI, this can be adapted with the participant being offered their own agent alongside a matched alternative, with contact and interaction time as the dependent measure over sessions. In this instance, care must be taken that the alternatives are matched as closely as possible, since an alternative option that is slower, less capable or visibly different, converts the test into a preference for competence, novelty, or aesthetics.

An alternative design would involve a \textit{substitution probe}. Here, multiple near-identical agents could be presented, with the only difference between them being their interaction history with the human. Identity (grounded in history of interactions) of the agents can be manipulated independently of capability and appearance. For instance, a study could include (i) the original agent with history, (ii) the original agent with history removed, (iii) the alternative agent with history (of original agent, transferred), and/or (iv) the alternative agent without history. With this template, selectivity can be separated from a preference for a ``character'' or persona, which can easily be reproduced in an alternative agent. Selectivity would predict a preference for the original agent, even when the history has been transferred. The probe has (inadvertently) been run before; this occurred after Sony's withdrawal of AIBO support in 2014, where owners opted to pay third-parties to keep their own pet robots running, rather than upgrading to newer models. Measurement is equally tractable, with a behavioural index adapted from pet-ownership research built from companion-robot activity logs \cite{takada2023}, and a scale developed with owners of the current model returning \emph{irreplaceability} as one of its five factors \cite{ichikura2024exploratory}. Although none of this was controlled, it establishes that the contrast is available in the field, and that people act on it at cost.%

Three published designs can serve as further inspiration. First, \cite{klumpe2025} contrasted a dog with a robotic replica over counterbalanced month-long phases with urinary oxytocin as the outcome. Further, \cite{mayer2026} ran a counterbalanced crossover of a robot against a human interviewer with a full endocrine panel, and  \cite{spatola2022precuneus} alternated a partner within subjects across 12 exchanges, fitting the response over time. We argue that all of these are promising directions insofar as they manipulated partner identity, although they all compared across \emph{kinds} of partner rather than between units of a specific kind (i.e. robot vs.  robot). Future studies can build on these studies, tightening the agent manipulation to allow this question to be answered.

There is also reason to think HRI is unusually well placed to do this. The hardest contrast in the study of social bonds is between a preference for \emph{this individual} and a preference for \emph{an individual of this kind, with this history}. In the comparative literature, contrast can only be approximated, because two animals cannot be made identical and an animal's interaction history cannot be extracted and installed in another. With an artificial partner, it can be run directly: an agent can be duplicated, matched exactly on capability and appearance, and its interaction history transferred, withheld or reassigned. The criterion that biology has had to approach obliquely for fifty years is, in the artificial case, a much more straightforward experiment.
 
\subsection{Question 3. What does the agent regulate?}
\label{sec:q3}

Question~3 looks to answer which, if any, of the physiological regulators Table \ref{tab:reg} an agent supplies. Unlike the other questions, the output of this question is more akin to an affective profile rather than a discrete ``yes/no'' answer. Since these regulatory channels disassociate (Section \ref{sec:magnitude}), it is possible for one or more of them to be present between a dyad. However, this should not be read as a cumulative sum of regulators (i.e. two regulators present do not necessarily mean a ``stronger bond'' than one), but rather different compositions. These compositions refer to the different affective profiles and role of the partner. For example, stress buffering and pain modulation are \textit{phasic}, since they require an activating condition, and regulate the affective and physiological cost of adversity \cite{hostinar2014,heinrichs2003}. The partner's presence is recruited once the body's prediction error has already occurred \cite{barrett2015interoceptive}. In contrast, synchrony and thermal--circadian co-regulation are \emph{tonic}: they can operate in ordinary, non-threatening contexts, and their regulation is the baseline that an individual holds in the presence of a partner. Here, the partner's presence has been written into the prediction, and their absence is a prediction error the body pays for \cite{beckes2011,coan2006,matthews2019, barrett2015interoceptive}. 

Building on the discussion in Section \ref{sec:bonding}, a study measuring these should consider therefore the following:
 
\begin{enumerate}[label=(\roman*),leftmargin=1.9em,itemsep=0.05em,topsep=0.2em]
\item each channel is explicitly measured by its own design, and reported independently;
\item at a minimum, comparison is agent-present against agent-absent. To align with Question ~2, a possible extension should also include a \textit{matched alternative present} condition;
\item an inanimate-agent control is included (where applicable to that regulator).
\end{enumerate}

This question is answered positively when at least one of the regulators is established against a suitable control condition (i.e. a matched alternative). 

\subsubsection{Potential Study Designs}
\label{subsec:q3_designs}

The standardised design to test the effect of \textit{social buffering} is the Trier Social Stress Test (TSST) \cite{allen2016tsst}. The test, and its child version, supply a validated stressor with characterised cortisol reactivity and recovery profiles. In this instance, stress buffering is potentiated in humans by oxytocin acting alongside social support \cite{heinrichs2003,hostinar2014}, and has been demonstrated intra- and cross-species \citet{beetz2012dogbuffer}. This paradigm can be adapted for HRI studies via longitudinal design, where participants engage with a known stressor (i.e. task, environmental, or social-based stressors) over time (e.g. once a week for a set period). In an ideal scenario, the buffering effect can be operationalised using salivary cortisol sampled at baseline and post-onset of stressor, scored as reactivity and recovery trajectory. \cite{pruessner2003auc}. However, considering that many HRI researchers do not work with wet lab assays,  electrodermal activity (EDA) and heart-rate variability (HRV) represent suitable alternatives. These would require continuous recording, measuring reactivity and recovery again, post-stressor onset. A buffering effect would be identified when reactivity is attenuated by the presence of the participants' agent (i.e. relative to being alone) and even relative to a matched alternative. Care must be taken to establish effects of both strerssor and agent(s)' presence individually, prior to their interaction. Further, this must ensure that any effects related to participant sex and the type of support (presence, vocal, physical contact, etc) provided by the agent are modelled, since these determine the sign of the valence of the effect as well as its amplitude \cite{kirschbaum1995sex,ditzen2007couple}. Components of this design have been run separately across HRI. For example, an inanimate-agent control in a cluster-randomised trial \cite{moyle2017paro}, a naturally occurring stressor has been combined with randomised availability \cite{beran2013vaccination}, a cortisol measure has been used with a contact manipulation that held for owners and non-owners alike \cite{imamura2023}, and a counterbalanced crossover design has been utilised \cite{mayer2026}. All o these studies can provide inspiration for HRI study designs.

For \emph{behavioural and physiological regulation}, a dyad can be recorded through a fixed, ordinary-use task---a timed conversation or a collaborative task---repeated across sessions in three conditions: with the participant's agent, with a matched alternative, and alone. The primary outcome here is tonic physiological state in the agent's presence (e.g. high-frequency HRV, respiratory sinus arrhythmia (RSA), EDA, or pupil diameter). Behavioural coupling---movement cross-correlation, speech-turn latency, gaze coupling, facial mimicry, gesture cadence---should be recorded alongside as a `mechanistic' secondary measure, as it indexes the route by which regulation is achieved. To isolate effects of the (shared) task structure, coupling between a particular human--robot dyad should be compared against a baseline, such as another participant's session with the same agent \citep{palumbo2017}. An agent whose output cadence is fixed by design can entrain a user without any dyadic events, with robot-led entrainment and mutual coordination both reported in HRI as ``synchrony'' \cite{jouaiti2024synchrony}. Thus, the agent's behavioural time series must be logged/entered as a candidate driver, by testing the direction of coupling, or more efficiently, by including a condition in which the agent runs a pre-recorded schedule.

\textit{Separation distress} is measurable at any interruption of the agent's availability, which can be included in an HRI study in a number of ways. The \textit{still-face} structure is a first example, in which an agent remains physically present and becomes briefly unresponsive before resuming  \cite{mesman2009stillface}. This is a well-validated developmental paradigm, which takes less than one minute, and is fully reversible within the same session. The second is a \emph{naturally occurring interruption}---for instance, a deprecation, migration, service incident, or a scheduled maintenance window \cite{defreitas2024,banks2024,zhang2026fragility,namvarpour2025}---with brief momentary assessment
either side.  The third alternative is a \emph{planned interruption} of the participant's agent. Such designs can be run concurrently in answering Question 2, where distress at losing a \textit{specific} agent can be established against loss of an alternative agent. A graded version (e.g. unresponsive, absent, permanently withdrawn) indicates which of the other regulators were in play; what returns on restoration is what was being regulated. However, one confound specific to artificial partners is that an unresponsive agent is also a malfunctioning agent. Thus, distress at a partner's withdrawal is not distinguishable from irritation at a broken tool, unless this is accounted for in the study design. 

\textit{Pain and threat modulation} is well-characterised in human dyads, where partner presence and affective touch attenuate subjective pain reports, autonomic arousal, and neural threat signatures \cite{vonmohr2018pain,coan2017handholding,pan2024touch}. Although HRI practitioners have successfully operationalised this channel using thermal pain, electrocutaneous threat anticipation, and targeted affective touch \cite{geva2020,yim2022,nakae2025}, scaling these paradigms across the broader field faces distinct practical bottlenecks. Ethically, deliberate use of noxious or threat-inducing stimuli restricts its routine application. Morphologically, contact-mediated paradigms require specialised, tactile-compliant hardware capable of naturalistic interactions such as handholding or affective stroking. To serve as a definitive index of bonding, future HRI deployments must combine these specialised embodiment designs with rigorous selectivity controls.

Establishing even a single selective regulator is sufficient to answer Question 3 affirmatively (Section~\ref{sec:q3}). A dyad demonstrating only stress buffering constitutes an affective social bond as equivalently as one exhibiting three distinct regulators. We do not interpret the presence of multiple regulators as a scalar measure of ``bond strength,'' but as a qualitative map of the agent's regulatory profile---one that predicts which physiological supports would fail upon the agent's withdrawal. Equally, the functional category of a regulator provides more diagnostic value than the overall regulator count. A strictly \emph{phasic} profile indicates a bond activated primarily by environmental adversity. Conversely, a profile containing \emph{tonic} regulators signifies a continuous affective arrangement, wherein the user's physiological baseline has structurally adapted to the agent's presence. We return to the latter point specifically in Section \ref{sec:ethics}.

\subsubsection{A Note on Methodological Feasibility }
We recognise that experimental designs of this type sit substantially outside the traditional methodological scope of HRI. Many HRI evaluations are conducted for short-term usability, algorithmic evaluation, and self-reported user experience, rather than longitudinal, physiological tracking of relationship formation. The paradigms we propose demand wet-lab assay handling (e.g., for salivary cortisol or oxytocin), continuous high-resolution psychophysiological recording, familiarisation periods measured in weeks, and rigorous ethical frameworks for safely inducing and relieving acute stress. Naturally, this infrastructure is rarely standard equipment in computer science or engineering laboratories.  We offer these protocols, in part, prospectively, as a blueprint for targeted interdisciplinary collaboration. By partnering with disciplines in which these paradigms are already native (e.g. developmental psychology, affective neuroscience, and psychoneuroendocrinology) HRI researchers can supply highly controllable, embodied artificial agents, while collaborators provide established measurement infrastructure and clinical safeguarding. In fact, a standing theoretical case for such cross-disciplinary integration in social robotics already exists \cite{cross2019socialbrains}. Embracing these shared methodological standards is the necessary next step for the field to transition from simply observing brief social interactions to empirically validating enduring socio-emotional bonds.

\subsection{Question 4. Are the security functions activated?}
\label{sec:q4}

Finally, Question~4 asks whether the agent provides the security functions ---safe haven and secure base --- in its presence. Grounded in foundational attachment theory, we treat a positive result here as a non-negotiable prerequisite for describing a human--robot tie as an ``attachment''. These functions are selective by their nature. A safe haven is not a source of comfort in general, but the support this person returns to. Similarly, a secure base is not \textit{any} available presence, but the specific one around which exploration is organised. A partner who could be exchanged for a matched stranger without consequence is not, on Ainsworth's second criterion, functioning as an attachment figure at all \cite{ainsworth1978,cimarelli2021}. Therefore, Question~4 presupposes Question~2. The selectivity contrast of Section~\ref{sec:q2} is a precondition of the design, and a study that has not established selectivity cannot return \emph{met} here, only \emph{not tested}. What Question~4 adds is the manipulation that makes a security function observable. Experimental designs must meet the same standard as those of the source disciplines. We therefore propose the following criteria in experimental design:

\begin{enumerate}[label=(\roman*),leftmargin=1.9em,itemsep=0.05em,topsep=0.2em]
\item an \emph{eliciting condition}: a stressor for the safe-haven function, or a novel or uncertain environment carrying some cost to explore for the secure-base function. 
\item the target's \emph{availability is experimentally manipulated} under that condition (present or withdrawn, or accessible or inaccessible)
\item an outcome measured \emph{under} the manipulation (i.e. by the measures admitted in Section~\ref{sec:q2}), showing that behaviour is organised with respect to the target's availability. In other words, exploration is sustained while the figure is available (and contracting when it is withdrawn) or a threat response is attenuated in its presence (but not it absence).
\end{enumerate}

\subsubsection{Potential Study Designs}
\label{subsec:q4_designs}

The cross-species study by \cite{zilchamano2012} and \cite{horn2013securebase} can offer inspiration for testing the secure-base function in HRI scenarios. After a familiarisation period with a robot (i.e. long enough to establish an interaction history), a participant can work on an open-ended or insoluble problem (e.g. a difficult manipulable puzzle \cite{gillet2020social} or a challenging game \cite{ravandi2025exploring}) for a fixed period across several sessions. The four conditions in \cite{horn2013securebase} can be replicated, with: (i) the participants' familiar agent absent, (ii) their agent present-but-silent, (iii) their agent present and encouraging (but not directly helping), and a (iv) matched alternative agent present and silent. A further condition may include a matched alternative agent present and encouraging.  The security function can be assessed broadly on task persistence over these conditions (i.e. time engaging with task,  latency to first disengagement, number of attempts, and resumption after failure). To account for effort rising in the presence of an encouraging other, and for reasons unrelated to security, further measures linked to checking-back behaviours (e.g. gaze to the agent, verbal references, latency to first check after a failure), should be included as secondary measures \cite{sroufewaters1977}. The secure base function would be seen when persistence rises in both own-agent conditions relative to absence, and does not rise with a matched alternative. In this paradigm, exploration is measured through a (sit-down) task, rather than physical movement through an environment, making it suitable for a range of agent embodiments. 

The study by \citet{gacsi2013safehaven} offers a template for assessing the \textit{safe haven} function. In that study, a threatening approach by a stranger produced a cardiac response, subsequently attenuated in the owner's presence. The threatening approach serves as the activating condition, and can be presented alongside a manipulation of the agent's availability---own agent available, matched alternative available, no agent---with cardiac and electrodermal measures recorded continuously. Although this resembles the stress buffering design of Question~3, the safe-haven function requires that an agent be \emph{sought} under threat, not (just) that arousal falls in its presence. Consequently, the primary outcomes are orientation and proximity-seeking toward the agent at activating condition, onset with attenuation of arousal as a secondary measure. An agent that lowers arousal (i.e. without being sought) has supplied physiological regulation (Question~3), rather than acting as a safe haven. In this design, any threat presented must be sufficient to activate the system, which is established in the no-agent condition; novelty and habituation must also be accounted for \cite{palmer2008,cimarelli2021}. Naturally occurring stressors, as seen in  \cite{beran2013vaccination,nozawa2026}, are an ethical route to this with vulnerable populations.%

Of the four questions in the procedure, we believe that Question~4 is the one whose designs are most immediately viable in HRI. The secure base design above requires a challenging task, 3-4 counterbalanced conditions, a fixed camera, and a means to record the time of various events (task persistence/engagement, latency, time to resumption, gaze time, etc). Every measure it uses is coded from video, using readily-available software. Unlike previous questions (e.g. Question 3) that may require physiological instrumentation, we propose that the entry-level protocol for this question is low. %

\begin{table*}[tp]
\centering
\small
\setlength{\tabcolsep}{5pt}
\caption{The classification provided by each set of answers to the procedure presented in Section~\ref{sec:classify}, resulting in one of six relationship classification. \checkmark\ established, --- not established, $\ast$ not required, $\cdot$ not reached.}
\label{tab:classify}
\begin{tabularx}{\textwidth}{c c c c Z}
\toprule
\textbf{Q1} & \textbf{Q2} & \textbf{Q3} & \textbf{Q4} & \textbf{Classification of the relationship} \\
\midrule
\checkmark & \checkmark & $\ast$ & \checkmark &
\textbf{Attachment.} Both selectivity and security regulation requirements are met. Answer to Q3 not required. \\
\addlinespace
\checkmark & \checkmark & \checkmark & --- &
\textbf{Affective social bond.} Both comparative requirements met, with the channel profile reported separately. \\
\addlinespace
\checkmark & \checkmark & --- & --- &
\textbf{Affiliative preference.} Ainsworth's affectional-bond criteria are met. However, regulation criterion not tested. \\
\addlinespace
$\ast$ & --- & \checkmark & --- &
\textbf{Non-specific regulation.} Physiological changes are measured with respect to robot's presence, without evidence that the state change is dependant on the specific agent \\
\addlinespace
\checkmark & --- & --- & --- &
\textbf{Affectional tie}. Ainsworth's affectional tie criteria met: Question~1 answered via observation. Selectivity of the agent and physiological regulation not tested. \\
\addlinespace
--- & $\cdot$ & $\cdot$ & $\cdot$ &
\textbf{No relationship to classify.} The affectional criteria are not met, usually because the encounter was too brief for persistence to be observable. The interaction may still be engaging, trusted and valued.\\
\bottomrule
\end{tabularx}
\end{table*}

\begin{table*}[tp]
\centering
\small
\setlength{\tabcolsep}{4pt}
\caption{The physiological regulators an affective social bond is composed of, grouped by when they operate.}
\label{tab:reg}
\begin{tabularx}{\textwidth}{Y{0.797} Y{1.007} Y{0.986} Y{0.944} Y{1.238} Y{1.028}}
\toprule
\textbf{Regulator} & \textbf{Physiological substrate} & \textbf{What an agent supplies} & \textbf{Observable proxy} & \textbf{Reference method} & \textbf{Practical proxy} \\
\midrule
\multicolumn{6}{@{}l@{}}{\textit{\textbf{Phasic} (Reactive/Event-Driven)}} \\
\addlinespace[0.2em]
Stress buffering &
HPA axis; cortisol response, CRF suppression \citep{hostinar2014} &
Contingent responsiveness &
Cortisol reactivity and recovery (HPA); EDA and HRV (autonomic) &
Trier Social Stress Test \citep{allen2016tsst,heinrichs2003}; salivary cortisol AUC$_i$ \citep{pruessner2003auc} &
Chest-strap ECG for HRV \citep{schaffarczyk2022polar}; wrist EDA \citep{milstein2020e4} \\
\addlinespace
Pain and threat modulation &
Endogenous $\mu$-opioid signalling \citep{machin2011,loseth2024} &
Presence; contact only if embodied &
Pain threshold, tolerance and rating &
Laser-evoked potentials and pain report \citep{vonmohr2018pain}; threat-of-shock handholding \citep{coan2017handholding} &
Contact against no-contact under thermal pain \citep{geva2020,nakae2025} \\
\addlinespace[0.5em]
\multicolumn{6}{@{}l@{}}{\textit{\textbf{Tonic} (Continuous/Baseline)}} \\
\addlinespace[0.2em]
Behavioural and physiological synchrony &
Interpersonal cardiorespiratory coupling \citep{goldstein2017coupling}; motor and speech timing \citep{ramseyer2011mea}; raised pain thresholds under synchrony \citep{cohen2010rowers} &
Turn-taking, prosody and gesture mirroring, by fixed policy &
Tonic autonomic state; movement, speech and cardiac cross-correlation &
Motion Energy Analysis \citep{ramseyer2011mea}; cross-recurrence \citep{coco2014crqa}; surrogate controls \citep{palumbo2017} &
Fixed-camera video; chest-strap ECG for cardiac coupling \citep{schaffarczyk2022polar}; the agent's own turn logs \\
\addlinespace
Thermal and circadian co-regulation &
Homeostatic regulation carried by contact \citep{hofer1994,fotopoulou2022} &
Requires contingent warmth and continuous presence &
Skin temperature; sleep timing and continuity &
Infrared thermography \citep{nazzari2024thermal,cardone2020thermalhri} &
Consumer actigraphy, for sleep timing and continuity\citep{yuan2024actigraphy} \\
\addlinespace[0.5em]
\multicolumn{6}{@{}l@{}}{\textit{\textbf{Disruption Signatures}}} \\
\addlinespace[0.2em]
Separation distress &
Opioid withdrawal \citep{panksepp1978,loseth2024}; HPA activation \citep{hostinar2014} &
Shutdown; persona update \citep{defreitas2024,banks2024} &
Negative affect, searching, protest, grief &
Consented interruption; still-face structure \citep{mesman2009stillface} &
Naturally occurring outage, with momentary assessment \citep{defreitas2024,banks2024} \\
\addlinespace
Selective aggression / exclusivity &
Vasopressin signalling \citep{winslow1993} &
Exclusivity to a unit or persona &
Resistance to substitution or sharing &
Response to reassignment; jealousy scale \citep{pfeiffer1989jealousy} &
Offer to transfer the agent's interaction history to another user \\
\bottomrule
\end{tabularx}
\end{table*}

 \subsection{Applying the procedure to published work}
\label{sec:apply}

To demonstrate the practical utility and diagnostic applicability of our procedure, we evaluate it against a representative sample of HRI studies on human--robot relationships, including those introduced in Section~\ref{sec:hriuse}. The results are summarised in Table~\ref{tab:apply}. For each study, we systematically apply each of the four questions, outlining the criteria applied and/or the measures used in the study, the term/vocabulary used by the authors in their description of a particular relationship, and the classification that our procedure would return. For simplicity, we establish whether Questions 1--4 were answered with relevant empirical evidence, rather than strictly enforcing our full proposed experimental design standards. Moreover, we do not consider this an exhaustive systematic sweep. Instead, this sample provides a representative cross-section of candidate studies, many discussed in this paper previously, and those that report relational, behavioural, or physiological data.

We note two observations from the table. First, reported terminology consistently diverges from the classification returned by our procedure, directly supporting the diagnostic mismatch identified by \citet{mitchell2025}. Many sampled papers claim a ``bond'' or ``social bond'' yet return a status of \emph{no relationship to classify}—not because they failed a specific diagnostic test, but because they lacked the baseline empirical conditions required for relational evaluation. Second, rather than collapsing studies along a weak-to-strong continuum, the procedure sorts the literature into four distinct, qualitative statuses with substantive differences between them. A single lab session and a four-year follow-up are not weak and strong instances of the same construct, nor is a study establishing selectivity without regulation equivalent to one establishing regulation without selectivity. This provides the concrete response to \citet{mitchell2025} outlined in Section~\ref{sec:magnitude}: rather than stretching a single term to accommodate weak ties, using distinct, criterial statuses keeps the full range of interactions in view while allowing researchers to make precise, determinate claims about each.

Interestingly, several studies that did not make any explicit claims about relational status nevertheless returned a determinate classification under our procedure. For instance, per-owner behavioural indices built from interaction logs \cite{takada2023} and surveys documenting grief following platform retirement \cite{defreitas2024,banks2024} satisfy Question 1 despite their authors asserting neither a bond nor an attachment. Similarly, the strongest evidence for persistence emerges from families retaining an agent unit four years after the conclusion of a six-month deployment \cite{zhao2025stayed}. Based on available data, these studies qualify as \emph{affectional ties} (Table~\ref{tab:classify}) and, with minor methodological additions, could have formally evaluated the presence of an \emph{attachment} or an \emph{affective social bond}.

The studies carrying the strongest physiological evidence \cite{wada2007paro,nozawa2026,imamura2023} and those carrying the strongest relational evidence \cite{zhao2025stayed,carpenter2016eod,defreitas2024} form entirely disjoint sets. Therefore, no dyad in the literature has both bonding criteria established. Partner selectivity is established explicitly in only one study \cite{knox2018aibo}. Other evaluations bearing on Question 2 either lack a comparative baseline \cite{takada2023,ichikura2024exploratory} or contrast across distinct categories rather than individual units \cite{spatola2022precuneus,klumpe2025,mayer2026}. For \emph{tonic} regulators, current platforms frequently generate temporal coordination as a byproduct of pre-programmed turn-taking. However, without surrogate controls, existing literature fails to support this coordination as dyadic, rather than passive entrainment to a fixed schedule \citep{palumbo2017,jouaiti2024synchrony}. Contact-mediated tonic channels—such as thermal, cardiac, and circadian co-regulation—remain entirely uninstantiated due to the hardware limitations detailed in Section~\ref{sec:q3}. For \emph{phasic} regulators, contingent responsiveness offers a plausible substrate for stress buffering, yet the lack of matched-alternative controls limits current findings to generalised, non-specific regulation. Paradoxically, the field is richest in data regarding the behavioural signatures of separation distress, leveraging commercial retirement events to track human responses to agent withdrawal at scale. Contemporary HRI thus possesses abundant data on what withdrawal reveals, but virtually none on the selective biobehavioural co-regulation being supplied.

\subsubsection{What a study that met the standard would look like}
\label{sec:benchmarks}

To supplement the application of our procedure above, we outline how several designs in our existing sample are already approaching positive resolution of a particular question. We propose these studies should be considered ``near misses'', and not failures outright. 

For \textbf{Question~2}, \citet{spatola2022precuneus} is structurally the closest existing design: the partner is alternated within subject, and the response is fitted over exchanges rather than compared at a single time point. Its current comparison runs across \emph{kinds} of partner---a human against a robot---rather than between units of one kind. Substituting a second, matched agent that differs from the first only in interaction history converts it into a selectivity test, and nothing else about the design has to change. The same is true of \citet{klumpe2025} and \citet{mayer2026}, both proposing counterbalanced, repeated sessions, with a physiological outcome. Changing the contrast to different robots, rather than robot against a dog or human would allow Question 2 to be answered in full.
 
For \textbf{Question~3}, both halves of the required design have been run separately. \citet{moyle2017paro} supplies an inanimate control, with behavioural and observational outcomes. What it currently lacks is a regulator. Conversely, \citet{beetz2012dogbuffer} supplies the regulator and a three-way contrast---a real dog, a friendly human, and a toy dog---with salivary cortisol as the outcome. A study combining the contrast of \citet{beetz2012dogbuffer}
with the control and the population of \citet{moyle2017paro} would answer Question~3 for a companion robot.
 
For \textbf{Question~4}, \citet{nozawa2026} comes closest. The study runs a true separation episode in the home, records cardiac measures across throughout, and logs the duration of physical contact with the robot during the episode, reporting that children ``significantly increased their physical interaction with the robot during separation''. That is the safe-haven outcome proper: the agent was sought under the activating condition. What the design lacks is the contrast. Availability is manipulated between families rather than within the dyad, so the comparison is between children who have a robot and children who do not, and no matched alternative unit is offered. Moving the manipulation inside the dyad and adding a second, matched robot would convert it into a safe-haven test in full.

However, we do not suggest that the authors of these respective papers have not been diligent, however. In almost every row, authors measured an authentic outcome and reported it fairly. However, what was lacking was shared vocabulary, by which a determinate partial result could be accurately reported. Instead, a partial result (i.e. where some of the criteria were met) acquired a whole-construct name. The classifications in Table~\ref{tab:classify}, along with the reporting standard provided in Section\ref{sec:classify}, offer a solution to such results.

\begin{sidewaystable*}[p]
\centering
\fontsize{6.5}{7.5}\selectfont
\setlength{\tabcolsep}{2pt}
\caption{The procedure applied to published human--robot and human--agent studies. \checkmark\ met; \ding{55}\ tested and not met; --- not tested. Rows are grouped by the status the reported evidence warrants. Definitions for each classification can be seen in Table \ref{tab:classify}}
\label{tab:apply}
\begin{tabularx}{\dimexpr\textheight-12pt\relax}{Y{0.700} Y{0.600} Y{1.275} c c c c Y{0.564} Y{1.861}}
\toprule
\textbf{Study} & \textbf{Term used} & \textbf{Criterion or measure used in study} & \textbf{Q1} & \textbf{Q2} & \textbf{Q3} & \textbf{Q4} & \textbf{Status returned} & \textbf{Notes} \\
\midrule
 
\multicolumn{9}{@{}l@{}}{\textit{\textbf{No relationship to classify}}}\\
\addlinespace[0.1em]
 
\citet{belpaeme2012} & ``social bonds'' (title; pp.~47--48) & Single forced-choice closeness item; 13 of 19 children completed three sessions &
--- & --- & --- & --- & No relationship to classify &
Labels of ``friend'', ``classmate'' and ``brother/sister'' were read as evidence of a social bond forming. \\
\addlinespace[0.1em]
 
\citet{salekshahrezaie2021} & ``bonding'' (abstract) & Single item (``I felt a bond with the Robot while it was speaking'') after two one-way speeches under three minutes each; $n=16$ &
--- & --- & --- & --- & No relationship to classify &
Two brief one-way exposures: too short for \textit{persistence} to be observable. \\
\addlinespace[0.1em]
 
\citet{xu2025care} & ``bonding'' (title) & Closeness, social attraction and desire for future interaction; hesitancy to replace the robot, effort on a task the robot requested &
--- & --- & --- & --- & No relationship to classify &
Single lab session, \textit{persistence} not tested.  \\
\addlinespace[0.1em]
 
\citet{kuhnlenz2020bonding} & ``social bonding'' (title) & Bonding \emph{induced} by small talk with emotional adaptation; outcomes were social presence, anthropomorphism and unsolicited helping &
--- & --- & --- & --- & No relationship to classify &
Bonding enters as the independent variable, applied in one session by a dialogue script. \\
\addlinespace[0.1em]
 
\citet{barakova2018} & ``bonding'' (title) & Engagement and empathy, adapted from an immersion questionnaire &
--- & --- & --- & --- & No relationship to classify &
Bonding operationalised as game engagement and empathy in a single session; across \textit{kind} comparison \\
\addlinespace[0.1em]
 
\citet{tanaka2007socialization} & ``bonding'' (abstract; conclusions) & Continuous audience response method: rated interaction quality across 45 sessions &
--- & --- & --- & --- & No relationship to classify &
\textit{Persistence} observed five months in situ. \textit{Separation distress} and \textit{emotional significance} not assessed. \\
\addlinespace[0.1em]
 
\citet{spatola2022precuneus} & ``social bonding'' (abstract) & Posterior cingulate response over twelve one-minute conversations per partner, partner alternated within subject &
--- & --- & --- & --- & No relationship to classify &
Alternating partner, but human vs. robot rather than robot vs. robot. \\
\addlinespace[0.1em]

\multicolumn{9}{@{}l@{}}{\textit{\textbf{Affectional tie}}}\\
\addlinespace[0.1em]
 
\citet{zhao2025stayed} & ``emotional attachment'' & Interviews at a four-year follow-up to a 180-day deployment; retention, placement and condition of each unit recorded &
\checkmark & --- & --- & --- & Affectional tie &
Persistence present: 18 of 19 families held the same unit four years later. No interruption occurred, so \textit{separation distress} is untested. \\
\addlinespace[0.1em]
 
\citet{sung2007roomba} & ``intimacy'' (abstract); ``attachment'' (body) & Ethnographic interview &
\checkmark & --- & --- & --- & Affectional tie &
Recorded across a mean ten months of ownership; no alternative unit was offered and no interruption occurred. \\
\addlinespace[0.1em]
 
\citet{carpenter2016eod} & ``attachment'' & Field interview; reported refusal of a functionally identical replacement &
\checkmark & --- & --- & --- & Affectional tie &
Refusal reported retrospectively from interview. \\
\addlinespace[0.1em]
 
\citet{defreitas2024} & ``emotional bonds'' & Survey and text analysis around discontinuation and persona-altering updates &
\checkmark & --- & --- & --- & Affectional tie &
Distress at an actual, unplanned interruption; no matched alternative to compare against.\\
\addlinespace[0.1em]
 
\citet{takada2023} & ``attachment'' (title; abstract) & Per-owner behavioural index (holding, calling the robot's name) from companion-robot activity logs &
\checkmark & --- & --- & --- & Affectional tie &
 \\
\addlinespace[0.25em]
 
\multicolumn{9}{@{}l@{}}{\textit{\textbf{Affiliative preference}}}\\
\addlinespace[0.1em]
 
\citet{knox2018aibo}; \citet{ichikura2024exploratory} & ``affectionate relationship''; ``bond'' & Interviews with owners and with a third-party repair firm; owner-derived scale returning \emph{irreplaceability}&
\checkmark & \checkmark & --- & --- & Affiliative preference &
Owners paid a third party to keep their own AIBO unit running rather than adopt a working replacement. \\
\addlinespace[0.25em]
 
\multicolumn{9}{@{}l@{}}{\textit{\textbf{Non-specific regulation}}}\\
\addlinespace[0.1em]
 
\citet{wada2007paro} & ``robot therapy'' & Interviews and social-network analysis; video of public areas; urinary analysis &
--- & --- & \checkmark & --- & Non-specific regulation &
Evaluated on a group sharing two units, not a single unit; the hormonal measures are stress markers.  \\
\addlinespace[0.1em]
 
\citet{nozawa2026} & ``bonding'', ``safe haven'' (title); ``attachment bond'' (abstract) & Cardiac measures and duration of contact with the robot across a separation--reunion episode; seven-day home deployment &
--- & --- & \checkmark & --- & Non-specific regulation &
Contact with the robot rose during separation. No matched alternative unit was offered, so specificity is not tested. \\
\addlinespace[0.1em]
 
\citet{imamura2023} & ``affiliative relationship'' (title; abstract) & Steady-state oxytocin, owners against non-owners; cortisol under a contact manipulation &
--- & --- & \checkmark & --- & Non-specific regulation &
Compares two groups of people; cortisol fell in owners and non-owners alike. The null on specificity is reported without an equivalence test. \\

\bottomrule
\end{tabularx}
\par\vspace{0.3em}
\end{sidewaystable*}

\subsection{The Cases of Embodiment and the Reciprocation Loop}
\label{sec:embodiment}
\label{sec:disembodied}
\label{sec:reciprocation}

Two properties of the artificial partner place an upper bound on the extent to which this procedure can be applied, and, importantly, the conclusions that we can draw from it. The first is the agent \textbf{embodiment}. Social robots need not be \textit{physically} embodied, and indeed, a large body of HRI studies are conducted with virtual agents. Users describe disembodied agents with the vocabulary of friendship, deepen in self-disclosure over months, and grieve their withdrawal \cite{skjuve2021,brandtzaeg2022,xiepentina2022,pentina2023,defreitas2024,banks2024,zhang2026fragility}; the robotics literature locates the advantage of physical robots in physical (and then social) presence rather than embodiment as such \cite{li2015presence}; here, physical embodiment raised social presence where contact was available, but produced null or negative effects once tactile interaction was restricted \cite{lee2006embodied}. However, insights from social cognition suggest a more nuanced role for physical presence. Because a physically embodied robot occupies a user's shared 3D peripersonal space, it grounds joint attention, spatial referencing, and sensorimotor resonance in ways that 2D screen agents cannot; this holds even in the absence of physical contact \cite{wainer2006embodiment,bainbridge2011presence,wykowska2016embodied}.  Embodiment therefore serves two functions.  First, it is a structural prerequisite for observing the contact-mediated channels in Table~\ref{tab:reg} (e.g. affective touch, thermal and cardiac co-regulation). Second, it acts as a cognitive and spatial amplifier for non-contact channels, such as behavioural synchrony and stress buffering.  %
Physical embodiment is not a precondition for the presence of an affective social bond under our procedure. Rather, it dictates the topography of the bond by limiting which regulatory channels can be observed (Question~2). If this account is correct, then we would predict that a disembodied agent, like a large language model, a virtual agent in 2D or 3D/VR space, and an embodied robot without a contact affordance should return the \emph{same} regulator profile. Similarly, an embodied agent supplying contingent contact should differ from these exclusively on the contact rows. %

The second property relates to \textbf{reciprocation} mechanisms. As we have mentioned previously (Section \ref{sec:regulation}), oxytocinergic mechanisms acting with dopamine consolidate the reward value of a \emph{particular} partner in biological systems \cite{insel1995ota,ross2009otr}. This is also implicated in the effects of social buffering \cite{heinrichs2003}, behavioural synchrony \cite{feldman2012}, and in other regulatory effects \cite{feldman2017}, reinforcing further social interactions and bond formation \cite{walum2018}. In biology, this runs in parallel, with each partner reinforcing the others', such that social interactions, selectivity, bonding, and its underlying regulation, become a feedback loop between two agents (see e.g cross-species example \cite{nagasawa2015}). However, in the human--robot case, the artificial agent emit the cues that engage a human's physiological regulatory system, but has itself no underlying regulatory circuitry that the interaction would alter (e.g. analogous to the human's homeostatic state/neuroendocrine system). Thus, the bidirectional affective feedback loop would remain open. Promising work such as \cite{maroto2024} is moving in this direction, developing a robot-side neuroendocrine model in which simulated oxytocin, vasopressin and dopamine modulate an autonomous robot's social behaviour as a function of its interaction history with a particular user. Whether such reciprocation and co-regulation is required in human--robot dyads, is consequential to the formation of a social bond or attachment, and what it may afford our bonding or attachment criteria above, all remain open questions.

\section{Ethical implications of studying attachment and social bonding in HRI}
\label{sec:design}
\label{sec:safety}
\label{sec:ethics}

In standard HRI discourse, the ethics of social robotics is frequently framed around \emph{deception}---the concern that users hold false beliefs about an agent's inner life, remediable through explicit disclosure. This remains a legitimate concern, but it is not most sharply raised in study of social relationships in human--robot dyads. The obligations that follow from positive relational verdict concerns two things instead: 1) what an established dependency costs the human when it is broken, and 2) differential risks associated with setting out to establish one. We address these in turn, outlining practical steps to mitigate risks where possible. Rather than offering definitive solutions, the discussion here is intended to highlight some of the considerations that HRI researchers should have in mind when designing and executing studies in human--robot relationships.

\subsection{Physiological and Psychological Dependency}

On the co-regulatory account developed in Section~\ref{sec:regulation}, the body budgets its regulatory effort on assumption that a particular partner will be available, treating proximity to that partner as its expected condition \citep{beckes2011,coan2006,matthews2019,schulkin2011}.
In this sense, dependency is not a metaphor for fondness, nor is it a cognitive attitude that a human can have towards a robot. It is a settled allocation of regulatory/homeostatic labour, and withdrawal of a partner to whom a person is bonded does not return that person to their initial baseline. Instead, it transfers a metabolic and physiological load back onto an individual whose set-points have already adjusted around the partner's availability. This clarifies why separation distress is disproportionate to a single instance of withheld comfort \citep{panksepp1978,loseth2024} . Because selectivity---the criterion distinguishing bonds from preferences for a class of object---is established by interaction history specific to one unit, that load cannot be redistributed by supplying another unit of the same model. Replacement is not mitigation once a tie is selective, and this is the point at which most assurances about continuity fail.

The regulator groups in Table~\ref{tab:reg} predict distinct ethical costs. \textit{Phasic} dependencies surface primarily under challenge, manifesting as altered stress responses when adversity arrives. Conversely, \textit{tonic} dependencies operate continuously across autonomic baselines, diurnal rhythms, and sleep architectures. Nothing in a participant's conscious day marks a tonic dependency, and standard deployment metrics (e.g.  engagement counters, usage logs, or self-report mood scales) will not register its presence or its interruption. A user may accurately report enjoying an agent without realising their baseline physiological arousal has come to depend on it. This makes the tonic case the ethically critical one, since it is simultaneously the most consequential to break and the least detectable using conventional HRI instruments. Uncovering (or avoiding) such dependencies requires tracking tonic bio-indicators across pre-exposure baselines and post-withdrawal observation windows. %

\subsection{Loss, Grief, and Disenfranchisement}

Involuntary separations are widespread across both commercial and research domains, occurring through platform deprecations, persona-altering model updates, server migrations, and hardware failures. The human consequences of these events are well-documented: users experience grief comparable to bereavement, mourning that far exceeds reactions to typical consumer product failures, and explicitly report the complete loss of a specific relational partner following persona shifts \citep{defreitas2024,banks2024,zhang2026fragility,namvarpour2025}. The structural profile of such loss lacks any biological analogue. When a platform or underlying model is retired, the separation is abrupt, absolute, unannounced, and experienced simultaneously across the entire user base. To contrast with a human context, this offers no preceding illness, no geographic distance, no gradual tapering of contact, and no possibility of reunion. While biological loss is typically attenuated by at least one of these factors, manufactured separations provide none. For researchers designing experimental interruptions, commercial deprecation events offer clear empirical proof that forced separation elicits severe physiological and affective disruption; importantly, this demands explicit mitigation strategies.

Crucially, this type of grief currently lacks social standing. Traditional bereavement is cushioned by societal recognition—cultural validation that the loss is real, the mourning is proportionate, and the mourner is entitled to support. Grieving an artificial agent receives no such validation, resulting in \emph{disenfranchised grief} \citep{doka2019disenfranchised}: a deep biobehavioural disruption that users are socially discouraged from expressing. Anticipating judgment, users may pre-emptively minimise their distress, obscuring it from view and preventing it from surfacing in post-deployment evaluations. Furthermore, these disruptions may ripple beyond the primary user. In fact, household members, partners, and care staff may absorb the hidden burdens of both dependency and eventual rupture, without having consented to either.

\subsection{The Risks of Inducing a Bond}

Evaluating Question~1 requires observing distress during a forced separation, transforming the protocol from passive observation into an active \emph{physiological intervention}. Because Question~1 demands empirical evidence of disruption upon interruption, scientific rigour and ethical cost scale together. A deployment long enough to establish persistence ($Q1$) and selectivity ($Q2$) is precisely the setting where induced withdrawal inflicts maximum homeostatic and psychological distress. Conversely, a protocol that carries negligible ethical cost is usually one that failed to establish a bond in the first place. This inherent tension requires researchers to select withdrawal designs with extreme care, adhering strictly to the tiered separation hierarchy outlined in Section~\ref{sec:q3}.

This methodological reality introduces three specific ethical challenges for experimental design. Firstly, the issue of \textit{incomplete enrolment consent}: consent obtained at baseline—before any tie exists—cannot cover an intervention that becomes consequential only after a biobehavioural dependency forms (i.e. over weeks of interaction). Participant consent must be treated as a dynamic, ongoing process that requires explicit re-verification prior to entering any separation phase. Secondly, we have the issue of \textit{measurement reactivity}. Administering attachment and bonding instruments is not a passive observational act. Repeatedly prompting participants to evaluate their relational tie encourages them to articulate and consolidate emotional commitments they might not otherwise have formulated. Thirdly, \textit{unmonitored study exits} present a similar challenge. Longitudinal HRI studies inevitably end; however, protocols frequently treat equipment collection as a routine administrative exit. In reality, an unbuffered removal at peak regulatory adjustment constitutes an unmonitored forced separation. Exit protocols and debriefing procedures must be integrated directly into the core study design.

Beyond experimental contexts, commercial architectures continuously optimise for retention and engagement—the exact empirical predictors of emotional reliance \citep{fang2025chatbot}. When adaptive agents (i.e. that increase accommodation as reliance grows \citep{chu2026}) are paired with unannounced persona shifts or sudden platform deprecations, they structurally replicate developmental \emph{fright without solution} \citep{mainhesse1990}. This represents a condition where a figure (serving as the source of security) becomes simultaneously the source of alarm. We recognise that, when read in reverse, our diagnostic criteria for detecting an attachment effectively double as an architectural blueprint for engineering one. While eliciting attachment is frequently cited as a target objective in consumer HRI \citep{yang2013,eum2026,szondy2024attachment}, inducing regulatory dependence (without provisioning for its long-term maintenance/safe withdrawal) is inherently harmful. Studies that seek to elicit such ties should explicitly declare which regulatory channels they intend to engage and what provisions are in place for eventual separation.

\subsection{Vulnerable Populations and Institutional Settings}

These ethical burdens compound significantly in standard HRI target populations, such as children, care-home residents, and older adults seeking assistance. These groups are simultaneously the most likely to establish regulatory dependencies, reliant on those regulators for daily functioning, and least equipped to decline or absorb induced withdrawal. Voluntariness is further compromised in institutional settings where deployments are assigned, leaving individuals poorly positioned to decline an agent's removal. Crucially, in young children, both the attachment hierarchy (the structured network of caregiving figures) and intrinsic self-regulatory mechanisms are still actively forming. Introducing artificial regulators into this developing network of extrafamilial social buffers \citep{gunnar2015,hostinar2015adolescence,doom2016} introduces unexamined risks to long-term socio-emotional development. For these reasons, we argue strongly against experimentally inducing partner separations in these populations, advising instead that research be limited to observing naturally occurring interruptions or non-vulnerable cohorts.

Furthermore, in institutional care settings, positive relational verdicts risk being ``weaponised'' during procurement. That is, evidence that residents form ties with robots can potentially be used to justify reductions in human caregiving staff/function. Explicitly reporting \emph{which specific regulators were established} ($Q3$), rather than claiming a generic "social bond," provides a vital safeguard against equating artificial regulation with human caregiving.

However, these safeguards do not constitute an argument against companion robotics. Non-human regulatory dependencies can be benign, competent adults retain autonomy to enter them, and the appropriate response to an established dependency is continuity, and not prohibition. Designing agents to route distress upward to human figures \citep{law2022,trinke1997hierarchies} ensures that security functions are supported without placing the system in a protective role it cannot fulfill.

\subsection{Regulatory Leverage and Limits}
\label{sec:regulatory}

Existing legal frameworks offer immediate leverage for operationalising these ethical obligations. Under GDPR Article 20, interaction history constitutes portable data—a provision directly relevant to human--artificial ties, as a shared interaction history underpins the selectivity criterion ($Q2$). Enforcing history portability thereby mitigates relational disruption during vendor migrations using an established legal right. Similarly, the EU AI Act's prohibition on exploiting vulnerabilities associated with age, disability, or specific socio-economic conditions (Article 5) currently lacks a standardised operational test. Importantly, an evidenced biobehavioural or regulatory dependency in a vulnerable population provides a concrete, candidate benchmark that our diagnostic procedure directly establishes. What remains absent across current frameworks is a mandate for relational continuity: no regulation obliges vendors to maintain a deprecated agent, provide advance notice proportionate to deployment duration, or preserve persona continuity across model updates. This represents an especially especially stark gap when contrasted with the strict lifecycle and de-servicing obligations enforced for medical devices with comparable dependency profiles.

Finally, two explicit boundaries define the scope of this classification framework. First, it does not assign moral duty: a dependency created by a vendor, borne by a user, and mediated by a deployer, distributes obligations across distinct actors with divergent interests. Second, it does not evaluate the agent's moral status. The procedure assesses a relationship strictly through its functional and physiological impact on the human user, granting no moral standing to the artifact itself. This boundary is most pronounced when contrasted with research that models attachment \emph{in} the agent itself \cite{canamero2006attachmentbonds,kaplan2001artificial}. In other words, where engineering an artificial system to instantiate an internal attachment behavioral system raises questions about the machine's own regulatory state—a dimension strictly beyond the scope of our diagnostic procedure.

\section{Conclusion}
\label{sec:conclusion}
Human--robot interaction has adopted the vocabulary of \emph{attachment} and \emph{bonding} faster than it has adopted the evidential standards that make those terms informative. Across the source traditions, the distinction is not in degrees, but of kind: an \emph{affective social bond} is evidenced by selective preference for a particular partner together with partner-specific physiological or behavioural \textit{regulation}, whereas \emph{attachment} requires that the partner functions as a safe haven under distress, and a secure base for exploration. Lacking experimental designs that systematically evaluate these distinct functions, HRI cannot determine whether it has observed an attachment, an affective social bond, or a weaker relational tie of a different kind—resulting in a proliferation of ungrounded definitions, non-convergent metrics, and fragmented evidential standards \cite{law2022,mitchell2025,rabb2022}. This paper has addressed this methodological deficit by establishing a four-question diagnostic classification procedure that links each biobehavioural criterion to validated experimental paradigms, specifying the minimum evidence required to warrant each relational label. The accompanying reference framework provides practical tools for both study design and standardised reporting. %

Our  procedure concerns classification and measurement, and does not, by itself, settle anything about whether human--artificial ties are beneficial or harmful for people. However, it make those downstream questions empirically tractable by forcing clarity about what kind of tie is present and what it consists in. The approach also makes the ethical claims more explicit: claiming or designing for attachments or social bonds in human--robot dyads incurs non-trivial responsibilities regarding user dependence, vulnerability, and the management of regulatory withdrawal. Because the criteria and paradigms adapted here originate from a select set of well-studied biological species, their application to human--robot dyads remains an empirical inference that the field must systematically test.

But HRI does not have long to resolve this. At a moment when managing artificial companionship alongside the preservation of human social bonds is being recognised well beyond this field \citep{kirk2025}, the experiments and reporting standards we call for are scientifically necessary and, we think, ethically urgent. We therefore ask the field to make a choice it has so far deferred: to adopt designs capable of warranting the claims of ``attachment'' and ``social bonding'', or to adopt alternative terminology that its measurements can actually support.

\backmatter

\section*{Statements and Declarations}

\textbf{Funding.} The authors declare that no funds, grants, or other support were received during the preparation of this manuscript.

\textbf{Competing interests.} The authors have no competing interests, financial or non-financial, to declare that are relevant to the content of this article.

\textbf{Ethics approval.} Not applicable.

\textbf{Author contributions.} I.K. conceived the paper and wrote the manuscript. E.B. reviewed the manuscript and
provided edits.

\bibliography{biblio}

\end{document}